\documentclass[conference]{IEEEtran}

\usepackage{floatflt}

\usepackage{color}
\usepackage{cite}

\usepackage[normalem]{ulem}
\usepackage{multirow}
\usepackage{graphicx}
\usepackage{amsmath}
\usepackage{amsthm}
\usepackage{amssymb}

\usepackage{verbatim}

\usepackage{psfrag,array}

\usepackage{subfigure}

\usepackage{stfloats}

\usepackage{amsmath}
\usepackage{epstopdf}
\usepackage{algorithmic}

\usepackage{booktabs}

\usepackage{url}

\newcommand{\p}[1]{\mathop{\mbox{\it p} } }

\renewcommand{\vec}[1]{\ensuremath{\boldsymbol{#1}}}
\newcommand{\be}{\begin{equation}}
\newcommand{\ee}{\end{equation}}
\newcommand{\ba}{\begin{array}}
\newcommand{\ea}{\end{array}}
\newcommand{\bea}{\begin{eqnarray}}
\newcommand{\eea}{\end{eqnarray}}
\newcommand{\bean}{\begin{eqnarray*}}
\newcommand{\eean}{\end{eqnarray*}}

\newcommand{\diag}{\mathop{\rm diag}}

\newcommand{\rmh}{^{\dag}}
\newcommand{\rmt}{^{\rm T}}

\newcolumntype{L}[1]{>{\raggedright\let\newline\\\arraybackslash\hspace{0pt}}m{#1}}
\newcolumntype{C}[1]{>{\centering\let\newline\\\arraybackslash\hspace{0pt}}m{#1}}
\newcolumntype{R}[1]{>{\raggedleft\let\newline\\\arraybackslash\hspace{0pt}}m{#1}}

\definecolor{white}{rgb}{1,1,1}

\begin{document}

\title{A Low-complexity Channel Shortening Receiver with Diversity Support for Evolved 2G Devices}

\author
{
\begin{tabular}{c}
Sha Hu$^{\dagger}$, Harald Kr\"{o}ll$^{\ddagger}$, Qiuting Huang$^{\ddagger}$, and Fredrik Rusek$^{\dagger}$ \\
$^{\dagger}$Department of Electrical and Information Technology, Lund University, Lund, Sweden \\
$^{\ddagger}$Integrated Systems Laboratory, ETH Z\"{u}rich, Switzerland\\
$^{\dagger}$\{sha.hu, fredrik.rusek\}@eit.lth.se, $^{\ddagger}$\{kroell, huang\}@iis.ee.ethz.ch\\
\end{tabular}
}

\maketitle

\begin{abstract}
The second generation (2G) cellular networks are the current workhorse for machine-to-machine (M2M) communications. Diversity in 2G devices can be present both in form of multiple receive branches and blind repetitions. In presence of diversity, intersymbol interference (ISI) equalization and co-channel interference (CCI) suppression are usually very complex. In this paper, we consider the improvements for 2G devices with receive diversity. We derive a low-complexity receiver based on a channel shortening filter, which allows to sum up all diversity branches to a single stream after filtering while keeping the full diversity gain. The summed up stream is subsequently processed by a single stream Max-log-MAP (MLM) equalizer. The channel shortening filter is designed to maximize the mutual information lower bound (MILB) with the Ungerboeck detection model. Its filter coefficients can be obtained mainly by means of discrete-Fourier transforms (DFTs). Compared with the state-of-art homomorphic (HOM) filtering based channel shortener which cooperates with a delayed-decision feedback MLM (DDF-MLM) equalizer, the proposed MILB channel shortener has superior performance. Moreover, the equalization complexity, in terms of real-valued multiplications, is decreased by a factor that equals the number of diversity branches.

\end{abstract}

\section{Introduction}

Although in some parts of the world 2G networks are replaced by cutting-edge 4G and beyond networks, nowadays 2G networks are the workhorse for emerging M2M communications\cite{m2m}, longevity of 2G networks is predicted not least only since it offers global ubiquitous coverage and has the largest number of subscribers worldwide.

In order to make 2G networks competitive with other standards which intend to conquer the 200kHz bands, several extensions have been proposed over time. Some extensions aim for better coverage while others aim for higher throughput and better robustness against interference. What many extensions have in common is the use of diversity. For example, in extended coverage GSM (EC-GSM), coverage is improved to reach, e.g., deep-indoor devices via blind repetition diversity\cite{ecgsm}. Hereby the same GSM radio burst is transmitted repeatedly to ease reception by the IoT device. Another extension called Evolved EDGE (E-EDGE)\cite{gsm} aims at improving the fallback capability of 2G networks, and to make user experience less frustrating when loosing broadband connectivity of e.g. 4G cellular networks. To provide a reliable fallback solution, E-EDGE offers up to 1Mbit/s downlink data rate. Moreover, antenna diversity is used to increase throughput under noisy and interference prone environments. This is especially attractive for 4G-enabled smart phones, where more that two receive antennas are specified.

Despite all the advantages that diversity offers regarding coverage extension and interference robustness, it makes signal processing in the receiver very complex, especially since single carrier 2G systems suffer from ISI and CCI, whereas the adjacent-channel interference can be mitigated through low-pass filters. To suppress CCI, the space-time filter (STF) is proposed in \cite{JP97}. It is a joint filtering and channel estimation algorithm that combines the multiple received streams into a single stream and maximizes the signal to interference and noise ratio. However, as shown in \cite{JP97}, the ISI should not be handled by the STF since it causes noise enhancement. Furthermore, after STF the ISI duration is prolonged, which increases the complexity for ISI equalization. On the other hand, due to the limited length of the training symbols, a joint optimization of the STF over multiple branches is infeasible. This is due to the fact that, the correlation matrix of the received samples corresponding to the training symbols becomes singular as the number of diversity branches increases.

Therefore, in this paper we propose a low-complexity receiver for ISI channels with multiple diversity branches, which is able to achieve the full diversity gain with a single stream equalizer. It contains two stages: (1) interference suppression to suppress the CCI, and (2) channel shortener to reduce the number of states in ISI equalization. Different from STF, in the first stage we suppress the interference based on the least square (LS) criteria and keep the ISI channel of the target user unchanged. Besides interference suppression, the receiver needs to deal with the ISI introduced by the frequency-selective channels.

The MLM equalizer \cite{KB90} implements the maximum log-likelihood sequence estimation\cite{F72} with soft decisions. However, the number of states in the equalizer increases exponentially with the ISI duration. Therefore, techniques of channel shortening were developed to reduce the number of the states in the equalizer by filtering the ISI channel with a prefilter. Traditionally, the minimal phase filter \cite{GOMB02} is utilized to concentrate the energy of the channel impulse response (CIR) to the first few taps. Efficient design of the prefiters based on homomorphic filtering can be found in \cite{BZKWH12,KZWR15}. The signal parts corresponding to the channel tails with smaller energy are removed from the received signal by the delayed-decision feedbacks~\cite{DC89} and results in the DDF-MLM equalizer.

In \cite{RP12}, the authors proposed a different design of the channel shortener and the MLM equalizer is based on the Ungerboeck model \cite{F15}. The channel shortener is designed to maximize the lower bound of the information rate corresponding to a mismatched detection model. We call this approach the MILB demodulator, which contains a MILB channel shortener and MLM equalizer with channel tails truncated. The MILB demodulator was successfully applied to E-EDGE system in~\cite{HK15}. However, \cite{RP12} and \cite{HK15} solely deal with a single receive antenna ISI channel.

In this paper, we extend the MILB demodulator to support diversity which combines multiple diversity branches into a single data stream. As shown in Section IV, the proposed two-stage receiver is efficient in suppressing the CCI. Furthermore, with significantly reduced complexity in the equalizer, the MILB demodulator has a superior performance than the homomorphic filtering based channel shortener followed by the DDF-MLM equalizer. We denote the latter approach as the HOM demodulator. 

\subsection*{Notations:}
Throughout the paper, $\vec{I}$ represents an identity matrix, superscripts ``\,$\rmt$\," and ``\,$\rmh$\," denote the matrix transpose and Hermitian transpose, respectively. In addition, ``\,$\mathrm{Tr(\,)}$\," is the trace operator, ``\,$\mathcal{R\{\,\}}$\," fetches the real part of a variable, ``\,$\ast$\," denotes the linear convolution, and ``\,$\otimes$\," is the tensor product. 

\section{Single-Input Multi-Output Signal Model}
Consider a single-input and multi-output (SIMO) system with $N$ diversity branches and $M$ interferers. The symbols are transmitted over frequency-selective channels with additive white noise. The received sample $y_k^n$ on the $n$th diversity branch can be modeled as
\bea \label{model1} y_k^n =\sum_{\ell=0}^{L\!-1}h_{\ell}^{n}x_{k-\ell}+\sum_{m=0}^{M-1}\sum_{\ell=0}^{\tilde{L}\!-1}p_{\ell}^{n,m}s_{k-\ell}^m+n_{k}^{n},\eea
where $x_{k}$ is the transmit signal of the target user, $s_{k}^m$ is the transmit signal from the $m$th interferer, and $n_{k}^{n}$ is the noise variable on the $n$th receive branch, all at time $k$. The $\ell$th tap of the CIR on the $n$th diversity branch corresponding to the target user and the $m$th interferer are denoted as $h_{\ell}^{n}$ and $p_{\ell}^{m,n}$, respectively.  The longest ISI duration over all branches of the target user and interferers are denoted as $L$ and $\tilde{L}$, respectively. We assume that the noise variables $n_{k}^{n}$ are zero-mean complex Gaussian random variables with variance $N_0$. 

Let $z_k^n$ denote the interference term on the $n$th branch as
\bea  z_k^n= \sum_{m=0}^{M-1}\sum_{\ell=0}^{\tilde{L}\!-1}p_{\ell}^{n,m}s_{k-\ell}^m+n_{k}^{n}, \notag \eea
then model (\ref{model1}) can be written as
\bea \label{model2}   y_k^n =\sum_{\ell=0}^{L\!-1}h_{\ell}^{n}x_{k-\ell}+z_k^n.\eea
Let $\vec{y}_k=[y_k^{0} \; y_k^{1} \;\ldots \;y_k^{N-1}]\rmt$ and $\vec{z}_{k}=[z_k^{0} \; z_k^{1} \;\ldots \;z_k^{N-1}]\rmt$ be the vectors that comprise the received samples and interference from all $N$ diversity branches, respectively. Then model (\ref{model2}) can be written in the vector form
\bea \label{model3}  \vec{y}_k =\sum_{\ell=0}^{L\!-1}\vec{h}_{\ell}x_{k-\ell}+\vec{z}_k,\eea
where the $N\times 1$ vector $\vec{h}_{\ell}$ comprising the $\ell$th tap CIR from all diversity branches reads
\bea \label{hl} \vec{h}_{\ell}= [h_{\ell}^{0} \; h_{\ell}^{1}\;\ldots \;h_{\ell}^{N}]\rmt. \eea
Further, define the $NK\times 1$ vectors $\vec{y}_{k:K}$ and $\vec{z}_{k:K}$, and the $(K+L-1)\times 1$ vector $\vec{x}_k$ as
{\setlength\arraycolsep{2pt}  \bea \vec{y}_{k:K}&=&[\vec{y}_k\rmt \; \vec{y}_{k+1}\rmt \;\ldots \;\vec{y}_{k+K-1}\rmt]\rmt, \notag \\
  \vec{z}_{k:K}&=&[\vec{z}_k\rmt \; \vec{z}_{k+1}\rmt \;\ldots \;\vec{z}_{k+K-1}\rmt]\rmt,  \notag \\
 \label{xk} \vec{x}_{k:K}&=&[x_{k-L+1} \; x_{k-L+2} \;\ldots \;x_{k+K-1}]\rmt. \eea}
\hspace{-2mm}Then the signal model (\ref{model3}) that comprises a total of $NK$ received scalar samples reads
\bea \label{model4} \vec{y}_{k:K}=\vec{H}\vec{x}_{k:K}+\vec{z}_{k:K},\eea
and the $NK\times (K+L-1)$ convolution matrix $\vec{H}$ is
{\setlength\arraycolsep{2pt} \bea \label{H} \vec{H}\!=\!\left[\!\begin{array}{ccccccc}  \vec{h}_{L-1}& \cdots&\vec{h}_{1}&\vec{h}_{0}&~&~&~\\ ~& \vec{h}_{L-1}&\ddots&\vec{h}_{1}&\vec{h}_{0}&~&~\\  ~&~&\ddots&\ddots&\ddots&\ddots&~ \\  ~&~&~&\vec{h}_{L-1}&\cdots&\vec{h}_1&\vec{h}_0\end{array} \!\right]\!. \notag\eea}
\hspace{-1.8mm}Based on signal models (\ref{model3}) and (\ref{model4}), in Section II-A below we lay down the channel estimation module that will be used in both stages of the receiver. In Section II-B we discuss the noise estimation that will be used in the second stage.

 \begin{figure*}[t]
\begin{center}
\scalebox{.44}{\includegraphics{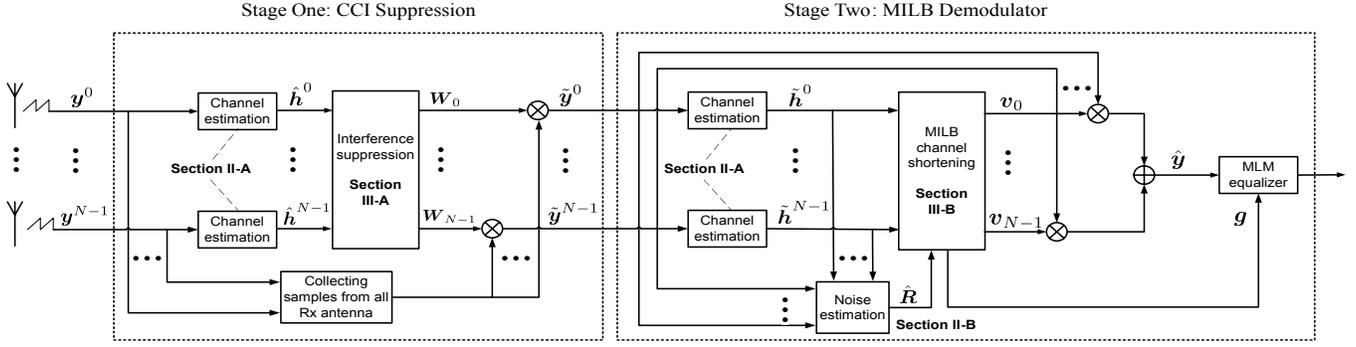}}
\vspace{-0mm}
\caption{The two-stage receiver with CCI suppression and the MILB demodulator. The branches $\boldsymbol{y}^0$ to $\boldsymbol{y}^{N-1}$ can alternatively be obtained from different blind repetitions in the repetition diversity \cite{ecgsm}.}
\label{fig:1}
\vspace{-3mm}
\end{center}
\end{figure*}

\subsection{Least Square Channel Estimation}
In an E-EDGE alike system, the CIR of the target user can be estimated through training symbols. Assume that $\vec{x}_{k_0:K_0}$ are the transmitted training symbols with length $\nu=K_0+L-1$ as defined in (\ref{xk}). From (\ref{model4}), it holds that
\bea  \label{model5} \vec{y}_{k_0:K_0}=\vec{H}\vec{x}_{k_0:K_0}+\vec{z}_{k_0:K_0}.\eea
Firstly, we define a useful operator $\mathcal{T}(\vec{a}, L)$ that generates a $(K-L+1)\times L$ matrix by cyclically shifting the elements of the $K\times 1$ vector $\vec{a}=[a_0 \; a_1 \;\ldots \;a_{K-1}]\rmt$ as below:
{\setlength\arraycolsep{2pt} \bea \label{T} \mathcal{T}(\vec{a}, L)\!=\!\left[\!\begin{array}{ccccccc}  a_{L-1}& a_{L-2}&\cdots&a_{0}\\a_{L}& a_{L-1}&\cdots&a_{1}\\ \vdots&\vdots&\vdots&\vdots \\  a_{K-1}& a_{K-2}&\cdots&a_{K-L}\end{array} \!\right]\!. \eea}
\hspace{-1.2mm}Define the $K_0\!\times\! L$ matrix $\vec{S}\!=\!\mathcal{T}(\vec{x}_{k_0:K_0}, L)$ which is generated from the $\nu\times 1$ training symbol vector $\vec{x}_{k_0:K_0}$. Further, let the vectors $\vec{y}_{k_0:K_0}^n$ and $\vec{z}_{k_0:K_0}^n$ comprise the the received samples and interference terms from time $k_0$ to $k_0\!+\!K_0\!-\!1$ on the $n$th diversity branch, respectively, which are defined as
{\setlength\arraycolsep{2pt} \bea \label{yn} \vec{y}_{k_0:K_0}^n&=&[ y_{k_0}^n\; y_{k_0+1}^n\;\cdots\;y_{k_0+K_0-1}^n]\rmt, \notag\\
 \vec{z}_{k_0:K_0}^n&=&[ z_{k_0}^n\; z_{k_0+1}^n\;\cdots\;z_{k_0+K_0-1}^n]\rmt.  \eea}
 \hspace{-1.5mm}Then from model (\ref{model5}), we have
{\setlength\arraycolsep{2pt} \bea \label{model6}  \vec{y}_{k_0:K_0}^n=\vec{S}\vec{h}^n+ \vec{z}_{k_0:K_0}^n, \eea}
\hspace{-3mm} where the $L\times 1$ vector $\vec{h}^n$ comprising the CIR on the $n$th diversity branch reads
\bea \vec{h}^n=[ h_{0}^n\; h_{1}^n\;\cdots\;h_{L-1}^n]\rmt.\notag\eea
 Therefore, the LS channel estimate of $\vec{h}^n$, which is denoted as $\hat{\vec{h}}^n$, can be obtained through
 \bea \label{LSH} \hat{\vec{h}}^n=\left(\vec{S}\rmh\vec{S}\right)^{-1}\vec{S}\rmh\vec{y}_{k_0:K_0}^n.\eea

\subsection{Noise Estimation}
Based on the channel estimates in (\ref{LSH}), the estimate of $\vec{h}_{\ell}$ in (\ref{hl}) can be obtained, which we denote as $\hat{\vec{h}}_{\ell}$. From model (\ref{model3}), the $N\times N$ covariance matrix $\vec{R}$ which represents the spatial correlation of the interference is estimated through
{\setlength\arraycolsep{2pt} \bea \label{R} \hat{\vec{R}}&=&\mathbb{E}\left[\hat{\vec{z}}_k\hat{\vec{z}}_k\rmh\right] \notag\\
&=&\frac{1}{K_0}\sum_{k=k_0}^{k_0+K_0-1}\hat{\vec{z}}_k\hat{\vec{z}}_k\rmh,\eea}
\hspace{-2mm} where
\bea\;\, \hat{\vec{z}}_k=\vec{y}_k-\sum_{\ell=0}^{L-1}\hat{\vec{h}}_{\ell}x_{k-\ell}.\notag\eea
Since the CCI is effectively suppressed by the interference suppression process on each diversity branch at the first stage, the temporary correlation of the interference is not further considered at the second stage in the receiver.

\section{The Structure of the Two-Stage Receiver}
In Section II, we have discussed the channel and noise estimates based on the training symbols. In this section, we elaborate on the structure of the two-stage receiver as depicted in Figure \ref{fig:1}. At the first stage, the samples are filtered by a group of filters $\vec{W}_{n}$ to suppress the CCI. The filtering is applied to mitigate interference without dealing with the ISI channel of the target user. Then at the second stage, based on the filtered samples $\tilde{\vec{y}}$, the channel estimate $\tilde{\vec{H}}$ and noise covariance matrix $\hat{\vec{R}}$ are updated and sent to the MILB channel shortening module, which generates the prefilters $\vec{v}_n$ and the coefficient $\vec{g}$ that will be used in the MLM equalizer. Samples on different branches are filtered and then summed up to a single data stream $\hat{\vec{y}}$. Unlike DDF-MLM equalizer in the HOM demodulator which needs parallel branch metric computation for each diversity branch, the MLM equalizer in the MILB demodulator only requires a single branch metric calculation. The complexity of the HOM and MILB demodulators will be further discussed in Section III-C.

\subsection{Stage One: CCI Suppression}

At the first stage, the received samples $\vec{y}^m$ are filtered by the filters $\vec{w}_{n,m}$ ($0\leq n,m<N$) to suppress the CCI. The filters have tap length $L_w$ (a design parameter) and are designed based on the signal model (\ref{model6}) to minimize the estimation error on the $m$th branch, which is
\bea  \label{e} \;\;\vec{e}_n\!=\!\min\limits_{\vec{w}_{n,m}}\left|\sum_{m=0}^{N-1}\vec{w}_{n,m}\!\ast\!\vec{y}_{k_0:K_0}^{m}\!-\!\hat{\vec{h}}^n\!\ast\!\vec{x}_{k_0:K_0-L_w+1}\right|^2\!\!. \notag \eea
The error $\vec{e}_n$ can be written in the matrix form as
{\setlength\arraycolsep{2pt}\bea  \label{ee} \vec{e}_n&=&\min\limits_{\vec{w}_{n,m}}\left|\sum_{m=0}^{N-1}\vec{T}_2^m\vec{w}_{n,m}-\vec{T}_3\hat{\vec{h}}^n\right|^2 \notag \\
&=&\min\limits_{\vec{W}_{n}}\left|\vec{\zeta}_2\vec{W}_{n}-\vec{T}_3\hat{\vec{h}}^n\right|^2,\eea}
\hspace{-1.5mm}where  
\bea \mathcal{\vec{\zeta}}_2=\big[\vec{T}_2^0\;\;\vec{T}_2^1\;\;\cdots\;\;\vec{T}_2^{N-1}\;\big], \notag \eea
 and
  \bea \vec{W}_n=\left[\vec{w}_{n,0}\rmt\;\vec{w}_{n,1}\rmt\;\cdots\;\vec{w}_{n,N-1}\rmt\right]\rmt.  \notag\eea
The $(K_0-L_w+1)\times L_w$ matrix $\vec{T}_2^m$ and $(K_0-L_w+1)\times L$ matrix $\vec{T}_3$ are generated by the operator $\mathcal{T}$ specified in (\ref{T}):
{\setlength\arraycolsep{2pt} \bea
\vec{T}_2^m&=&\mathcal{T}(\vec{y}_{k_0:K_0}^m, L_w), \notag \\
\vec{T}_3&=&\mathcal{T}(\vec{x}_{k_0:K_0-L_w+1}, L),\notag \eea}
\hspace{-2.5mm} where $\vec{y}_{k_0:K_0}^m$ and $\vec{x}_{k_0:K_0-L_w+1}$ are the vectors comprising of received samples on the $m$th diversity branch and the training symbols as in (\ref{yn}) and (\ref{xk}), respectively. The estimate $\hat{\vec{h}}^n$ on the $n$th branch has been obtained through (\ref{LSH}). Taking the first order differential of $\vec{W}_n$ in (\ref{ee}) yields the optimal filter
\bea \label{wn} \vec{W}_{n}=\mathcal{\vec{\zeta}}_2\rmh\left(\mathcal{\vec{\zeta}}_2\mathcal{\vec{\zeta}}_2\rmh\right)^{-1}\vec{T}_3\hat{\vec{h}}^n. \notag\eea
The received samples $\vec{y}^m$ are then filtered by the filters $\vec{w}_{n,m}$ to obtain the purified data on $n$th diversity branch as
\bea \label{wy} \tilde{\vec{y}}_n=\sum_{m=0}^{N-1}\vec{w}_{n,m}\vec{y}^m.\notag\eea
After filtering, the signal model (\ref{model4}) still holds for a data transmit block with size $K+L-1$ as
\bea \label{model8} \tilde{\vec{y}}_{k:K}=\vec{H}\vec{x}_{k:K}+\tilde{\vec{z}}_{k:K},\eea
where $\tilde{\vec{z}}_k$ is the residual interference after the CCI suppression. Furthermore, updated channel and noise estimates are obtained through (\ref{LSH}) and (\ref{R}) with filtered samples $\tilde{\vec{y}}_n$ in (\ref{model8}), which will be sent into the MILB channel shortening module in the second stage of the receiver.

\subsection{Stage Two: MILB Channel Shortening Algorithm}
The MILB channel shortener extends the framework developed in \cite{RP12} to deal with multiple diversity branches. With ISI channels, the MILB channel shortening is developed when the block size $K+L-1$ is infinite, in which case, we can let the channel matrix $\vec{H}$ represent circular convolution instead of the linear convolution with a $L$-tap ISI channel on each diversity branch. This approximation\footnote{Under the case that $L\!-\!1$ zero-padding tail symbols are inserted between successive transmit blocks in order to eliminate interblock interference, the linear convolution is equivalent to the circular convolution.} can reach any given precision \cite{W88} and is a result of Szeg\"{o}'s eigenvalue distribution theorem \cite{Sze}. With such assumption, we can rewrite the model (\ref{model8}) as, 
\bea \label{model9} \vec{Y}=\vec{H}\vec{X}+\vec{Z},\eea
where $\vec{Y}$ and $\vec{X}$ are $K\times 1$ vectors representing the received samples and transmit signal, respectively. The vector $\vec{Z}$ is a complex Gaussian noise vector that obeys $\vec{Z}\sim \mathrm{CN}(\vec{0} ,\vec{I}\otimes \vec{R})$. The $NK\times K$ block circulant matrix $\vec{H}$ is generated by its first column $[ \vec{h}_{0}\;\vec{h}_{1}\;\cdots\;\vec{h}_{L-1}\;\vec{0}\;\cdots\;\vec{0}]\rmt$.
Assuming that the demodulator operates on the Ungerboeck model $T(\vec{Y}|\vec{X})$, which is
\bea \label{model7} T(\vec{Y}|\vec{X}) = \exp\left(-2\mathcal{R}\left\{\vec{X}\rmh\vec{V}\rmh\vec{Y}\right\}+\vec{X}\rmh\vec{G}\vec{X}\right),\eea
the lower bound of the information rate is defined as
\be  \label{metric} I_{\mathrm{R}} = -\mathfrak{h}(\vec{Y})+\mathfrak{h}(\vec{Y}|\vec{X}),\ee
where $\mathfrak{h}$ is the entropy operator. Following the same approach as in \cite{RP12}, a closed form for $I_{\mathrm{R}} $ in (\ref{metric}) can be reached
\bea \label{GMI} I_{\mathrm{R}} =K+\log \big(\det(\vec{I}+\vec{G})\big)-\mathrm{Tr}\big(\vec{B}(\vec{I}+\vec{G})\big),\eea
with the optimal $K\times K$ prefilter matrix $\vec{V}$ reads
\bea \label{wopt} \vec{V}_{\mathrm{opt}}=\left(\vec{H}\vec{H}\rmh+\vec{I}\otimes \vec{R}\right)^{-1}\vec{H}\left(\vec{I}+\vec{G}\right),\eea
where $\vec{B}$ is the mean square error matrix
\bea  \label{mse} \vec{B}=\left(\vec{H}\rmh\left(\vec{I}\otimes \vec{R}^{-1}\right)\vec{H}+\vec{I}\right)^{-1}.\eea
The operation $\vec{V}\rmh\vec{Y}$ filters each diversity branch separately and the samples after filtering are summed up as
 \bea\label{ty} \hat{\vec{y}}=\vec{V}\rmh\vec{Y}=\left(\sum_{n=0}^{N-1} \vec{v}_n\ast\vec{y}_n\right)\rmt.\eea
In (\ref{model7}), for the purpose of channel shortening, the matrix $\vec{G}$ is constrained to be a band-shaped Hermitian Toeplitz matrix with a memory length $\nu<L$, that is, only the middle $2\nu+1$ diagonals can take non-zero values. We define the first $\nu+1$ non-zero elements in the first column of $\vec{G}$ as
 \bea \label{g} \vec{g}=[g_0\;g_1\;\cdots\;g_{\nu}]. \notag\eea
Furthermore, $\vec{I}+\vec{G}$ is constrained to be positive definite\cite{RP12} and we assume that
\bea\label{matG}\vec{I}+\vec{G}=\vec{U}\vec{U}\rmh,\eea
where $\vec{U}$ is a $K\times K$ upper triangular Toeplitz matrix generated from the vector $\vec{u}$, which comprises the first $\nu+1$ non-zero elements in the first row of $\vec{U}$ as
\bea \vec{u}=[u_0\;u_1\;\cdots\;u_{\nu}]. \notag\eea
Denote $\check{\vec{u}}=[u_{\nu-1}^{\ast}\;\cdots\;u_0^{\ast}]$, and from (\ref{matG}) it holds that,
\bea \label{uu} [g_{\nu}\;\cdots\;g_{1}\; g_0+1\;g_1^{\ast}\;\cdots\;g_{\nu}^{\ast}]=\vec{u}\ast\check{\vec{u}}. \eea
The optimal $\vec{u}$ is designed to maximize $ I_{\mathrm{R}}$ in (\ref{GMI}), and the optimal prefilter $\vec{v}_n$ and coefficient $\vec{g}$ can be solved through (\ref{wopt}) and (\ref{uu}), respectively. The algorithm is based on the DFT and inverse DFT (IDFT) operations and is summarized in Algorithm-1. The derivation is provided in Appendix A.

\begin{center}
 \begin{tabular}[t]{ m{22em}  }
  \toprule
   \label{Algorithm-1}
Algorithm-1: MILB channel shortening. \\
 \hline
 \;
 Input: Channel estimate $\hat{\vec{H}}$ and correlation matrix $\vec{R}$;\\
\quad 1.  DFT of $\vec{h}_\ell$ for all diversity branches to obtain \\
\qquad\!  frequency response $\vec{\lambda}_{k}$ in eq. (\ref{Fh});\\ 
\quad 2.  Calculate $\Delta_k\!=\!\left(\vec{\lambda}_k\rmh \vec{R}^{-1}\vec{\lambda}_k\!+\!1\right)^{\!\!-1}$ in eq. (\ref{deltak});\\ 
\quad 3.  IDFT of $\Delta_k$ to obtain $b_s$, which are elements\\
\qquad\! of MSE matrix $\vec{B}$ in eq. (\ref{bm});\\ 
\quad 4.  Calculate the optimal vector $\vec{u}$ based on $b_s$ as \\
\qquad\! in eq. (\ref{u0}) and eq. (\ref{u00});\\
\quad 5.  Calculate the optimal $\vec{g}$ based on eq. (\ref{uu});\\
\quad 6.  DFT of the optimal $\vec{u}$ to obtain $U_m$ in eq. (\ref{Um});\\ 
\quad 7.  Calculate $\vec{\Theta}_s\!=\! \frac{|U_s|^2\vec{R}^{-1}\vec{\lambda}_s}{1+\vec{\lambda}_s\rmh\vec{R}^{-1}\vec{\lambda}_s}$ in eq. (\ref{thetam});\\
\quad 8.  IDFT of $\vec{\Theta}_s$ to obtain prefilters $\vec{v}_n$ in eq. (\ref{vn});\\ 
Output: $\{\vec{v}_n\}_{n=0}^{N-1}$ and $\vec{g}$.\\
 \hline
\end{tabular}
\end{center}


%

\subsection{Complexity Analysis}
It was shown in\cite{HK15} that, computing the channel shortening prefilter coefficients in the MILB demodulator requires half the complexity as in the HOM demodulator. With diversity branches, the savings can also be achieved with Algorithm-1. This is because that, the inversion of covariance matrix $\vec{R}$ is required in both demodulators\footnote{In the HOM demodulator, the colored noise needs be whitened prior to the homomorphic filtering on each diversity branch.}, and step 2 and 7 in Algorithm-1 require a low amount of scalar multiplications and inversions compared to the case with a single diversity branch in\cite{HK15}. 

Next we evaluate the complexity of equalizers in the HOM and MILB demodulators, which is measured by the the number of real multiplications (one complex multiplication is counted as four real multiplications) per symbol stage. Notice that, the HOM demodulator requires a memory storage and updating process of the feedback symbols in all states with DDF-MLM equalizer. However, since the MILB channel shortener has truncated the channel tails, the MLM equalizer requires no feedback and is a simpler process than DDF-MLM equalizer.


Below we assume that the memory length in both equalizers is $\nu$ and the cardinality of the symbol modulation alphabet is $\mathcal{S}$. Then the number of states for both equalizers equals
\bea N_{\#} = \mathcal{S}^{\nu}. \notag \eea 
The branch metric calculation of the DDF-MLM equalizer in the HOM demodulator is based on the Forney model, and at the $k$th stage is calculated as
{\setlength\arraycolsep{1pt}  \bea &&\gamma\left( x_k,\cdots, x_{k-\nu}| \hat{x}_{k-\nu-1},\cdots, \hat{x}_{k-L+1}\right)\propto\notag \\
&&\qquad \qquad\qquad\quad \sum_{n=0}^{N-1}\left|\hat{y}_k^n-\sum_{\ell=0}^{\nu}h_{\ell}^n x_{k-\ell}-\sum_{\ell=\nu+1}^{L-1}h_{\ell}^n\hat{x}_{k-\ell}\right|^2\!, \notag\eea}
\hspace{-2.2mm}where $( x_k,\cdots, x_{k-\nu})$ are the symbols deduced from the current state, $\hat{y}_k^n$ is the filtered data obtained in (\ref{ty}), and $(\hat{x}_{k-\nu-1},\cdots, \hat{x}_{k-L+1})$ are the feedback symbols on the survival path leading to the current state. Hence, it needs $N(4L+2)$ real multiplications to calculate $\gamma$. Since the branching factor is $\mathcal{S}$, the total complexity at each stage for the DDF-MLM equalizer is
\bea C_{\mathrm{HOM}}=N(4L+2)N_{\#}\mathcal{S}=N(4L+2)\mathcal{S}^{\nu+1}. \eea
On the other hand, the Ungerboeck model based branch metric of the MLM equalizer in the MILB demodulator is
\bea \gamma(x_k,\cdots, x_{k-\nu})\!\propto\!-2\mathcal{R}\left\{\!x_k^{\ast}\!\left(\hat{y}_k^n\!-\!\sum_{\ell=1}^{\nu}g_\ell x_{k-\ell}\!\right)\!\right\}\!+\!g_0 |x_k|^2.\notag\eea
Since $g_0 |x_k|^2$ is the same for all $K$ stages, it can be pre-calculated\footnote{When $x$ is modulated with a constant amplitude such as $M$-PSK, this term can be removed from the calculation.} and the complexity of this part can be ignored. Hence, it needs $(4\nu+2)$ real multiplications to calculate $\gamma$. Since the branching factor is also $\mathcal{S}$, the total complexity at each process stage is
\bea C_{\mathrm{MILB}}=(4\nu+2)N_{\#}\mathcal{S}=(4\nu+2)\mathcal{S}^{\nu+1}. \eea
To obtain low complexity, 
we set $\nu=1$, and it holds that
\bea \frac{C_{\mathrm{MILB}}}{C_{\mathrm{HOM}}}=\frac{(4\nu+2)\mathcal{S}^{\nu+1}}{N(4L+2)\mathcal{S}^{\nu+1}}=\frac{3}{N(2L+1)}.\notag\eea
For different modulations, the number of real multiplication is listed in Table 2 where we assume $N=2$ and $L=8$.

\begin{center}
Table 2. Real Multiplication Number per Stage in Equalizer.
\vspace{-2mm}
 \begin{tabular}[t]{ |c|c|c|c|c|c |} 
 \hline
  Modulation   & $S$&$N_{\#}$ & $\nu$&$C_{\mathrm{HOM}}$&$C_{\mathrm{MILB}}$ \\
  \hline
  GMSK  & 2&2 & 1&320&24 \\
   \hline
  8PSK  & 8&8 & 1&5120&384 \\
   \hline
  16QAM & 16&16 & 1&20480&1536 \\
   \hline
  32QAM &32&32 & 1&81920&6144 \\
 \hline
\end{tabular}
\end{center}
\vspace{4mm}

On the other hand, instead of computing $h_{\ell}^n{x}_{k-\ell}$ and $g_{\ell}^n{x}_{k-\ell}$ directly at each stage, when symbols ${x}_{\ell}$ utilize a fixed alphabet (which does not hold for GMSK modulation), they are the same for all $K$ stages. Hence, they can be pre-calculated and stored. Then the branch metric calculation can use look-up-table (LUT) operations. With such an approach, the number of real multiplications required to calculate one branch metric is $N\mathcal{S}$ and $\mathcal{S}$ for the HOM and MILB demodulators, respectively. Therefore, $C_{\mathrm{MILB}}$ is still decreased by a factor of $N$ over $C_{\mathrm{HOM}}$. Meanwhile, the number of LUT operation in the MILB demodulator is only $(\nu+1)/L$ of the number of LUT operations in the HOM demodulator.

\section{numerical results}
In this section we evaluate the performance of the proposed low-complexity two-stage receiver. The uncoded bit error rate (BER) and block error rate (BLER) are measured. We assume an E-EDGE system with a single transmit antenna and two receive antennas ($N=2$). The uncoded BER and BLER are measured with different modulation and coding schemes (MCSs), and tested under typical urban (TU) and hilly terrain (HT) channels. The channel profiles are set according to the 3GPP specification \cite{3GPP1}. We consider four different MCSs that are specified in Table 3 and $\nu=1$ in all tests. In the plots, we denote the signal to noise power ratio as $S/N$ and the signal to CCI power ratio as $S/I$, both in dB, respectively.

\begin{center}
Table 3. MCS for E-EDGE Downlink \cite{3GPP2}.
\vspace{-2mm}
 \begin{tabular}[t]{|c|c|C{2cm}|C{1.5cm}|} 
 \hline
  MCS   & Modulation&User Data Rate (kbps) & Coding Rate\\
     \hline
  MCS1  & GMSK&8.8 & 0.53 \\
   \hline
  MCS5  & 8PSK&22.4 & 0.37 \\
   \hline
 MCS8 & 16QAM&54.4 & 0.67 \\
   \hline
   MCS10 &32QAM&67.2 &0.65 \\
 \hline
\end{tabular}
\end{center}
\vspace{4mm}

\begin{figure}
\begin{center}
\scalebox{.26}{\includegraphics{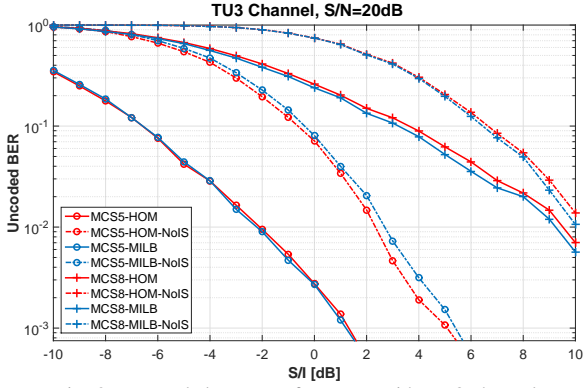}}
\vspace*{-15mm}
\caption{\label{fig:2} Uncoded BER performance with TU3 channel.}
\vspace*{-6mm}
\end{center}
\end{figure}

\begin{figure}
\begin{center}
\scalebox{.26}{\includegraphics{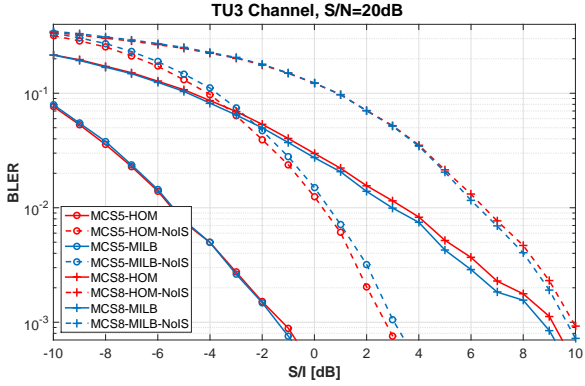}}
\vspace*{-15mm}
\caption{\label{fig:3} BLER performance with TU3 channel.}
\vspace*{-6mm}
\end{center}
\end{figure}

In Figure \ref{fig:2} and Figure \ref{fig:3}, we evaluate the CCI suppression performance for the MILB and HOM demodulators under TU channel at speed 3km/h. The $S/N$ is fixed to 20dB and the performance without interference suppression is denoted with suffix \lq{}NoIS\rq{}. As shown in Figure \ref{fig:2} and Figure \ref{fig:3}, both the uncoded BER and BLER are significantly boosted by suppressing the CCI. At $10\%$ uncoded BER, the gain with CCI suppression is around 6dB for MCS5 and 3dB for MCS8. The gain of MCS8 is reduced since, when $S/I$ increases the interference level decreases, hence the impact of CCI suppression also degrades. As shown in Table 2, although the complexity of the MILB demodulator is much less than the HOM demodulator, with CCI suppressing the performance is almost the same with MCS5 as shown in both figures. With MCS8, the MILB demodulator shows a gain of about 0.2dB at 10\% BLER above the HOM demodulator.

\begin{figure}
\vspace*{-6mm}
\begin{center}
\scalebox{.286}{\includegraphics{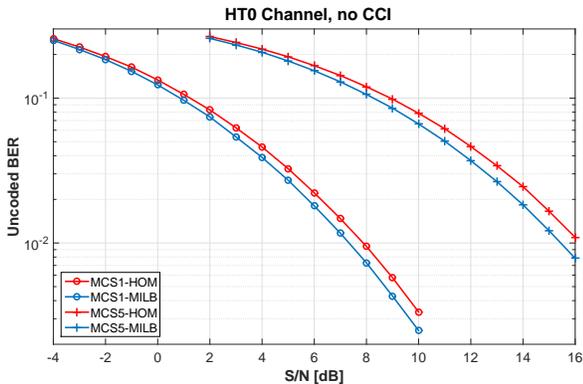}}
\vspace*{-8mm}
\caption{\label{fig:4} Uncoded BER performance with HT0 channel.}
\vspace*{-6mm}
\end{center}
\end{figure}

\begin{figure}
\vspace*{-0mm}
\begin{center}
\hspace{-0mm}
\scalebox{.286}{\includegraphics{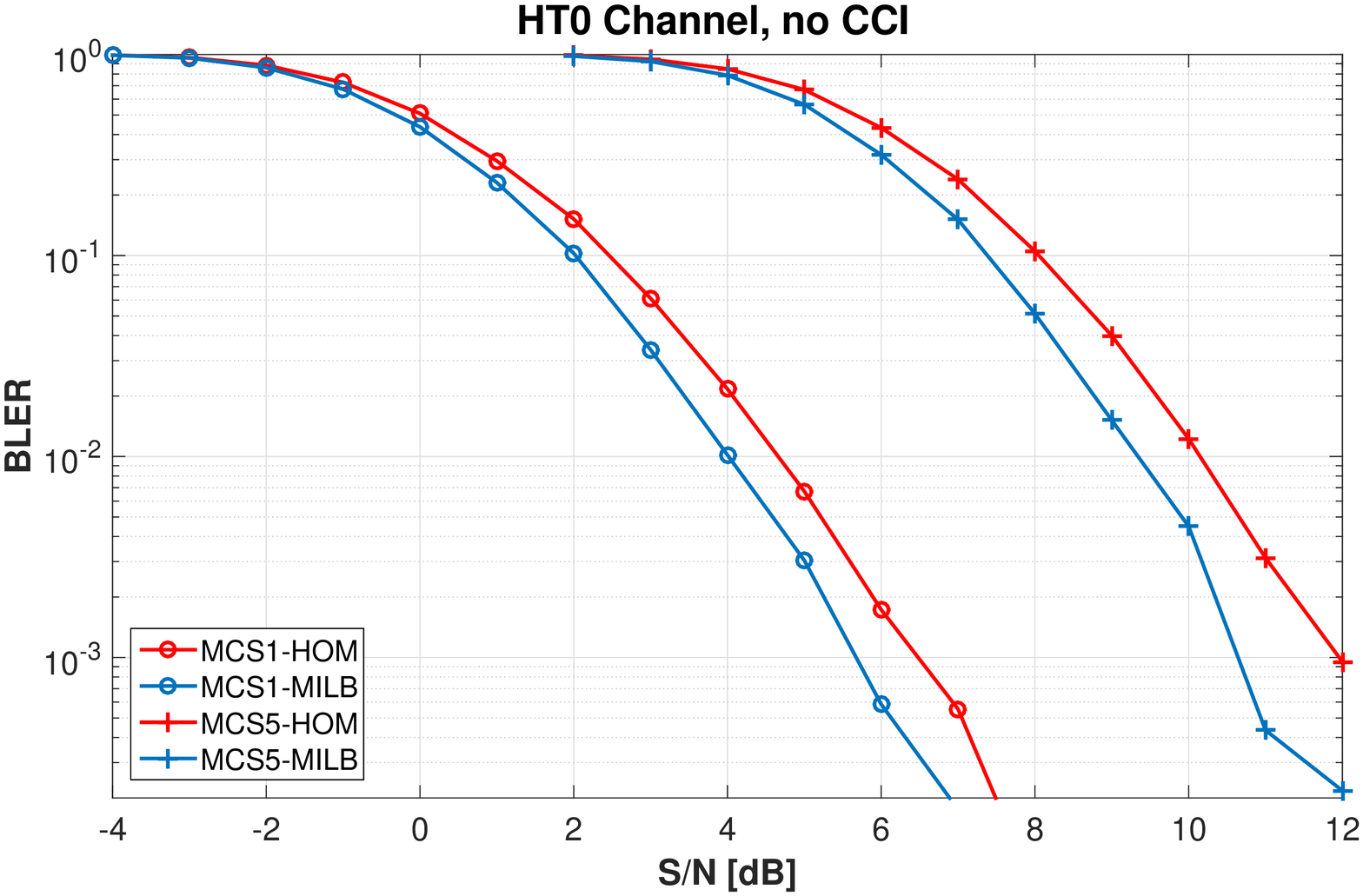}}
\vspace*{-8mm}
\caption{\label{fig:5} BLER performance with HT0 channel.}
\vspace*{-5mm}
\end{center}
\end{figure}

\begin{figure}
\vspace*{-0mm}
\begin{center}
\scalebox{.26}{\includegraphics{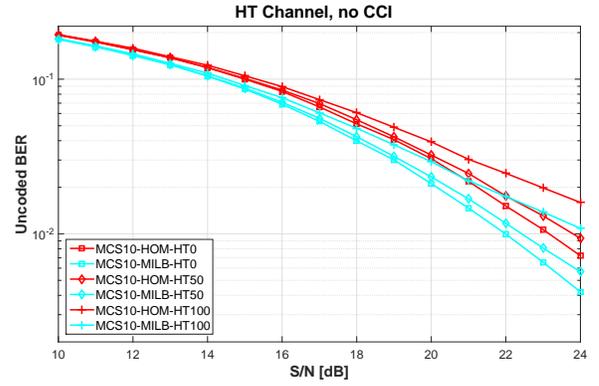}}
\vspace*{-15mm}
\caption{\label{fig:6} Uncoded BER performance with HT0, 50 and 100 channels.}
\vspace*{-5mm}
\end{center}
\end{figure}

\begin{figure}
\vspace*{-6.5mm}
\begin{center}
\scalebox{.26}{\includegraphics{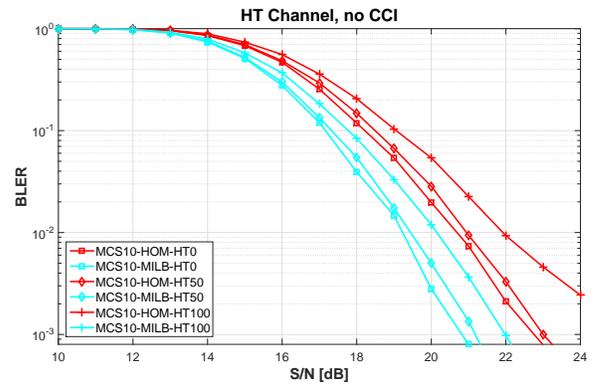}}
\vspace*{-15mm}
\caption{\label{fig:7} BLER performance with HT0, 50 and 100 channels.}
\vspace*{-6mm}
\end{center}
\end{figure}

Next we test the cases without CCI and under HT channel, which has long tails and is difficult for channel shortening. As shown in Figure \ref{fig:4} and Figure \ref{fig:5}, the MILB demodulator is superior to the HOM demodulator both with MCS1 and MCS5 under HT channel at no speed. With MCS10, as shown in Figure \ref{fig:6} and Figure \ref{fig:7}, the MILB demodulator is around 1dB better at $10\%$ uncoded BER and 2dB better at $1\%$ BLER than the HOM demodulator at all speeds. Moreover, HT channel at 100 km/h is around 1dB worse than HT channel at no speed at $1\%$ BLER. 

\section{Conclusions}
In this paper we proposed a low complexity two-stage receiver with interference suppression and MILB channel shortening for single carrier SIMO systems. The first stage is CCI suppression, which removes the co-channel interference. In the second stage, the MILB channel shortener generates both the optimal prefilters and coefficient for the Ungerboeck model based MLM equalizer. The received samples from different diversity branches are summed up to a single data stream prior to the MLM equalizer. The proposed receiver is evaluated for an E-EDGE system. As shown by the numerical results, the first stage is very effective in suppressing the CCI. In the second stage, although the MILB demodulator truncates the channel tails and cooperates with a reduced-state MLM equalizer, it shows superior performance compared with the traditional homomorphic filter which cooperates with the DDF-MLM equalizer. Furthermore, the MILB channel shortening algorithm uses the discrete-Fourier transforms, which allows reusing resources from e.g. OFDM modems.

\section*{Appendix A}
Denote the $K\times K$ Fourier matrix as $\vec{F}$ and the $(s,n)$th element as $F(s,n)=\exp\left( -2j\pi s n/K\right)$. The block circulant matrix $\vec{H}$ has the decomposition
\bea \label{FH} \vec{H}=(\vec{F}^{-1}\otimes\vec{I})\vec{\Lambda}\vec{F},\eea
where $\vec{\Lambda}=\diag(\vec{\lambda}_0\;,\vec{\lambda}_1,\;\ldots\;,\vec{\lambda}_{K-1})$ is block diagonal and each block element $\vec{\lambda}_k=[\lambda_{k,0}\;\lambda_{k,1}\;\cdots\;\lambda_{k,N-1}]\rmt$ is the $N\times 1$ column vector that comprises the DFT
 \bea \label{Fh} \lambda_{k,n}=\sum_{\ell=0}^{L-1}h_ {\ell}^{n}\exp\left( -\frac{2j\pi k\ell}{K}\right).\eea
Inserting (\ref{FH}) back into (\ref{mse}), $\vec{B}$ can also be decomposed as
\bea \label{deB} \vec{B}=\vec{F}^{-1}\vec{\Delta}\vec{F}, \notag\eea
where the $K\times K$ diagonal matrix $\vec{\Delta}$ is
\bea \vec{\Delta}=\left(\vec{\Lambda}\rmh\left(\vec{I}\otimes \vec{R}^{-1}\right)\vec{\Lambda}+\vec{I}\right)^{-1}, \notag\eea
and the $k$th diagonal element of $\vec{\Delta}$ reads
 \bea  \label{deltak} \Delta_k=\left(\vec{\lambda}_k\rmh \vec{R}^{-1}\vec{\lambda}_k+1\right)^{-1}. \eea
Next we define the IDFT
  \bea \label{bm} b_s=\frac{1}{K}\sum_{k=0}^{K-1}\Delta_k\exp\left( \frac{2j\pi ks}{K}\right), \; 0\leq s <K.\eea
By \cite[Proposition 2]{RP12}, the optimal $\vec{U}$ that maximizes ({\ref{GMI}}) is given by
\bea \label{u0} u_0=\frac{1}{\sqrt{b_0-\vec{b}_{\nu}(\vec{B}_{\nu})^{-1}\vec{b}_{\nu}\rmh}}, \eea
and
\bea \label{u00}[u_1\; u_2\;\cdots\; u_{\nu}] =-u_0\vec{b}_{\nu}(\vec{B}_{\nu})^{-1},\eea
where $\vec{b}_{\nu}$ is defined as
 \bea \vec{b}_{\nu}= [b_1^{\ast}\;b_2^{\ast}\;\cdots\;b_{\nu}^{\ast}]. \notag \eea
and the $\nu\times \nu$ sub-matrix $\vec{B}_{\nu}$ of $\vec{B}$ is
 {\setlength\arraycolsep{2pt} \bea \vec{B}_{\nu}&=&\left[\begin{array}{cccc}  b_0& b_1^{\ast}&\cdots&b_{\nu-1}^{\ast}\\b_1& b_0&\cdots&b_{\nu-2}^{\ast}\\ \vdots&\vdots&\vdots&\vdots \\ b_{\nu-1}& b_{\nu-2}&\cdots&b_0\end{array} \right]\!. \notag \eea}
\hspace*{-1.5mm}Hence, the optimal vector $\vec{g}$ defined in (\ref{g}) can be obtained through the optimal $\vec{u}$ as in (\ref{uu}). Defining the DFT
\bea \label{Um} U_s=\sum_{k=0}^{\nu}u_k^{\ast}\exp\left( -\frac{2j\pi ks}{K}\right), \; 0\leq s <K,\eea
and denoting $\vec{U}$ as the $K\times K$ diagonal matrix with $U_m$ being its $m$th diagonal element, then the optimal $\vec{V}_{\mathrm{opt}}$ in (\ref{wopt} ) can be decomposed as
\bea \label{FV} \vec{V}_{\mathrm{opt}}=\vec{F}^{-1}\vec{\Theta}(\vec{F}\otimes\vec{I}), \notag \eea
where the $K\times K$ block diagonal matrix $\vec{\Theta}$ is calculated by
\bea \vec{\Theta}=\left(\vec{\Lambda}\vec{\Lambda}\rmh+\vec{I}\otimes\vec{R}\right)^{-1}\!\vec{\Lambda}\vec{U}\vec{U}\rmh, \notag\eea
and the $m$th block element $\vec{\Theta}_s=[\Theta_{s,0}\;\Theta_{s,1}\;\cdots\;\Theta_{s,N-1}]$ is an $N\times 1$ vector that can be calculated as
 \bea \label{thetam}  \vec{\Theta}_s= |U_s|^2\left(\vec{\lambda}_s\vec{\lambda}_s\rmh+ \vec{R}\right)^{-1}\vec{\lambda}_s = \frac{|U_s|^2\vec{R}^{-1}\vec{\lambda}_s}{1+\vec{\lambda}_s\rmh\vec{R}^{-1}\vec{\lambda}_s}. \eea
Finally, the optimal filter $\vec{v}_n$ ($0\leq n <N$) for $n$th diversity branch is obtained through the IDFT as
 \bea \label{vn} \vec{v}_n=\frac{1}{K}\sum_{s=0}^{K-1} \Theta_{s,n}\exp\left( \frac{2j\pi sn}{K}\right).\eea

\bibliographystyle{IEEEtran}

\end{document}